\input harvmac
\input labeldefs.tmp
\writedefs
\overfullrule=0pt

\input epsf
\def\fig#1#2#3{
\xdef#1{\the\figno}
\writedef{#1\leftbracket \the\figno}
\nobreak
\par\begingroup\parindent=0pt\leftskip=1cm\rightskip=1cm\parindent=0pt
\baselineskip=11pt
\midinsert
\centerline{#3}
\vskip 12pt
{\bf Fig. \the\figno:} #2\par
\endinsert\endgroup\par
\goodbreak
\global\advance\figno by1
}
\newwrite\tfile\global\newcount\tabno \global\tabno=1
\def\tab#1#2#3{
\xdef#1{\the\tabno}
\writedef{#1\leftbracket \the\tabno}
\nobreak
\par\begingroup\parindent=0pt\leftskip=1cm\rightskip=1cm\parindent=0pt
\baselineskip=11pt
\midinsert
\centerline{#3}
\vskip 12pt
{\bf Tab. \the\tabno:} #2\par
\endinsert\endgroup\par
\goodbreak
\global\advance\tabno by1
}
\def\der{\partial}
\def\d{{\rm d}}
\def\e#1{{\rm e}^{#1}}
\def\E#1{{\rm e}^{\textstyle #1}}%

%

\def\Gam{{\mit\Gamma}}

\def\Th{Thistlethwaite}
%
\def\pre#1{ (preprint {\tt #1})}
%
%
%
\lref\HTW{J. Hoste, M. Thistlethwaite and J. Weeks, 
{\sl The First 1,701,936 Knots}, {\it The Mathematical Intelligencer}
20 (1998) 33--48.}
\lref\Ozanam{J. Ozanam {\sl R\'ecr\'eations math\'ematiques et physiques}
(Paris, 1694, 2 vols.). [Revised by Montucla (Paris, 1778, 4 vols.);
English translation by Hutton (London, 1803, 4 vols.)]. We here cite
p. 222, vol. 4 of the 1725 edition. We do not know whether the idea was
already manifest in the books by C. Bachet (1612 and 1624), and by
C. Mydorge (1630).}
\lref\Leibniz{G. W. von Leibniz, cited in
{\sl Histoire de l'Acad\'emie royale des Sciences de Paris}
pour l'ann\'ee 1771, p. 55.}
\lref\Vandermonde{A.-T. Vandermonde, {\sl Remarques sur les probl\`emes de
situation},
M\'em. de l'Ac. des Sc. de Paris pour l'ann\'ee 1771, p. 566 (Paris, 1774).}
\lref\AV{I.Ya. Arefeva and I.V. Volovich, 
{\sl Knots and Matrix Models}, {\it Infinite Dim.
Anal. Quantum Prob.} 1 (1998) 1\pre{hep-th/9706146}).}
\lref\BIPZ{E. Br{\'e}zin, C. Itzykson, G. Parisi and J.-B. Zuber, 
{\sl Planar Diagrams}, {\it Commun. Math. Phys.} 59 (1978) 35--51.}
\lref\BIZ{D. Bessis, C. Itzykson and J.-B. Zuber, 
{\sl Quantum Field Theory Techniques in Graphical Enumeration},
{\it Adv. Appl. Math.} 1 (1980) 109--157.}
\lref\DFGZJ{P. Di Francesco, P. Ginsparg and J. Zinn-Justin, 
{\sl 2D Gravity and Random Matrices}, {\it Phys. Rep.} 254 (1995)
1--133.}
\lref\tH{G. 't Hooft, 
{\sl A Planar Diagram Theory for Strong 
Interactions}, {\it Nucl. Phys.} B 72 (1974) 461--473.}
\lref\MTh{W.W. Menasco and M.B. \Th, 
{\sl The Tait Flyping Conjecture}, {\it Bull. Amer. Math. Soc.} 25
(1991) 403--412; 
{\sl The Classification of Alternating 
Links}, {\it Ann. Math.} 138 (1993) 113--171.}
\lref\Ro{D. Rolfsen, {\sl Knots and Links}, Publish or Perish, Berkeley 1976.}
\lref\STh{C. Sundberg and M. Thistlethwaite, 
{\sl The rate of Growth of the Number of Prime Alternating Links and 
Tangles}, {\it Pac. J. Math.} 182 (1998) 329--358.}
\lref\Tutte{W.T. Tutte, {\sl A Census of Planar Maps}, 
{\it Can. J. Math.} 15 (1963) 249--271.}
\lref\Zv{A. Zvonkin, {\sl Matrix Integrals and Map Enumeration: An Accessible
Introduction},
{\it Math. Comp. Modelling} 26 (1997) 281--304.}
\lref\KM{V.A.~Kazakov and A.A.~Migdal, {\sl Recent progress in the
theory of non-critical strings}, {\it Nucl. Phys.} B 311 (1988)
171--190.}
\lref\KP{V.A.~Kazakov and P.~Zinn-Justin, {\sl Two-Matrix Model with
$ABAB$ Interaction}, {\it Nucl. Phys.} B 546 (1999) 647
\pre{hep-th/9808043}.}
\lref\ZJZ{P.~Zinn-Justin, J.-B.~Zuber,
{\sl Matrix Integrals and the Counting of Tangles and Links},
proceedings of the 11th 
International Conference on Formal Power Series and Algebraic 
Combinatorics, Barcelona June 1999; to appear in
{\it Discrete Mathematics}\pre{math-ph/9904019}.}
\lref\ZJZb{P.~Zinn-Justin and J.-B.~Zuber, {\sl On the Counting of Colored
Tangles}, {\it Journal of Knot Theory and its Ramifications} 
9 (2000) 1127--1141\pre{math-ph/0002020}.}
\lref\PZJ{P.~Zinn-Justin, {\sl Some Matrix Integrals
related to Knots and Links}, proceedings
of the 1999 semester of the MSRI ``Random Matrices
and their Applications'', MSRI Publications Vol. 40 
(2001)\pre{math-ph/9910010}.}
\lref\PZJb{P.~Zinn-Justin, {\sl The Six-Vertex Model on
Random Lattices}, {\it Europhys. Lett.} 50 (2000) 15--21\pre{cond-mat/9909250}.}
\lref\IK{I.~Kostov, {\sl Exact solution of the Six-Vertex
Model on a Random Lattice}, {\it Nucl. Phys.} B 575 (2000) 
513--534\pre{hep-th/9911023}.}
\lref\KAUF{L.H.~Kauffman, {\sl Knots and Physics},
World Scientific Pub Co (1994).}
\lref\KAUFb{L.H.~Kauffman, {\sl Virtual Knot Theory}\pre{math.GT/9811028}.}
\lref\JZJ{J.~L.~Jacobsen and P.~Zinn-Justin,
{\sl A Transfer Matrix approach to the Enumeration of Knots}\pre{math-ph/0102015}.}
\lref\JZJb{J.~L.~Jacobsen and P.~Zinn-Justin,
{\sl A Transfer Matrix approach to the
Enumeration of Colored Links}\pre{math-ph/0104009}.}
\lref\EK{B.~Eynard and C.~Kristjansen, {\sl More on the exact solution of the
$O(n)$ model on a random lattice and an investigation of the case $|n|>2$},
{\it Nucl. Phys.} B 466 (1996) 463--487\pre{hep-th/9512052}.}
\lref\KoS{I.K.~Kostov, {\it Mod. Phys. Lett.} A4 (1989), 217\semi
M.~Gaudin and I.K.~Kostov, {\it Phys. Lett.} B220 (1989), 200\semi
I.K.~Kostov and M.~Staudacher, {\it Nucl. Phys.} B384 (1992), 459.}
\lref\KPZ{V.~G.~Knizhnik, A.~M.~Polyakov and A.~B.~Zamolodchikov,
{\sl Fractal structure of 2D quantum gravity},
{\it Mod.~Phys.~Lett.~A} 3, 819--826 (1988);
F.~David,
{\sl Conformal field theories coupled to 2D gravity in the conformal gauge},
{\it Mod.~Phys.~Lett.~A} 3, 1651--1656 (1988);
J.~Distler and H.~Kawai,
{\sl Conformal field theory and 2D quantum gravity},
{\it Nucl.~Phys.} B 321, 509 (1989).}
\lref\PZJc{P.~Zinn-Justin, {\sl The General $O(n)$ Quartic Matrix Model and
its Application to counting Tangles and Links}\pre{math-ph/0106005}.}
\Title{
\vbox{\baselineskip12pt\hbox{\tt math-ph/0111011}}}
{{\vbox {
\vskip-10mm
\centerline{The Combinatorics of Alternating Tangles:}
\vskip2pt
\centerline{from theory to computerized enumeration}
}}}
\medskip
\centerline{J.~L.~Jacobsen {\it and} P.~Zinn-Justin}\medskip
\centerline{\sl Laboratoire de Physique Th\'eorique et Mod\`eles 
Statistiques}
\centerline{\sl Universit\'e Paris-Sud, B\^atiment 100}
\centerline{\sl 91405 Orsay Cedex, France}
\vskip .2in
\noindent 

We study the enumeration of alternating links and tangles, considered up 
to topological (flype) equivalences. A weight $n$ is given to each connected
component, and in particular the limit $n\to 0$ yields information about
(alternating) knots. Using a finite renormalization scheme for an 
associated
matrix model, we first reduce the task to that of enumerating planar
tetravalent diagrams with two types of vertices (self-intersections and
tangencies), where now the subtle issue of topological equivalences has 
been eliminated. The number of such diagrams with $p$ vertices scales as 
$12^p$ for
$p\to\infty$. We next show how to efficiently enumerate these diagrams (in
time $\sim 2.7^p$) by using a transfer matrix method. We give results
for various generating functions up to 22 crossings.
We then comment on their large-order
asymptotic behavior.

\Date{10/2000}

\newsec{Introduction}

The mathematical description of knots and related combinatorial
objects is a subject with a history of several centuries. Giving an
account of the progress made over the years is beyond the scope of the
present paper, and we refer to \HTW\ for an introduction to the existing
literature (from the 19th century onwards). An old idea, first
published by Ozanam \Ozanam\ in 1694, is to treat a knot as a planar
diagram with over- and underpasses. This trick, however simple it may 
appear,
is essential to the present work, since it allows us to effectively 
reduce the
dimension of the problem from three to two, and in particular to apply the
powerful tools of matrix models in the planar limit.
The classification of such diagrams with respect to topological
equivalences calls for a particular geometrical framework (tagged
{\it geometria situs} by Leibniz \Leibniz), as was probably first pointed
out by Vandermonde \Vandermonde\ in 1771.


For technical reasons, we shall here constrain our attention to the subset
of {\it alternating knots} (and various generalizations such as links and
tangles), which by definition possess a representation as a diagram in 
which
over- and underpasses alternate. The first efforts in this domain date 
from
the late 19th century through the pioneering works of Tait, Kirkman and
Little. Tait originally thought that all knots were alternating, but 
Little
made it evident (though no proof was known at that time) that 
non-alternating
knots start appearing at order $p=8$ intersections. It is now believed 
that
alternating knots are asymptotically subdominant (the corresponding 
statement
has been proven rigorously in the case of links \STh). However, the 
enumeration of
alternating knots remains a very appealing mathematical problem, and to 
this
date no explicit formula giving the number of knots at order $p$ is known.

The subtle question of factoring out topological equivalences for 
alternating
knots is facilitated by two remarkable theorems which were first 
conjectured
by Tait, but only proved rigorously quite recently. First, alternating
diagrams can be reduced to diagrams having minimal crossing number by
iteratively eliminating irrelevant crossings through the move shown in 
Fig.~\nprime~a
\KAUF. Second, two such {\it reduced} alternating diagrams are
topologically equivalent if and only if they are connected by a sequence 
of
simple transformations (see Fig.~\flyp) known as {\it flypes} \MTh.

In the first part of the paper we shall show that counting alternating 
knots
modulo the flype equivalence is equivalent to computing the generating
function for a certain class of planar tetravalent diagrams that are
illustrated in Fig.~2. These diagrams possess two types of vertices
(self-intersections and tangencies), which are weighted separately. The 
proof
is inspired by the renormalization procedure originally invented within 
the
context of quantum field theory (QFT), and here applied to the random matrix 
model
that generates the above-mentioned planar diagrams. In particular we show 
that
factoring out topological equivalences is tantamount to including various
counterterms in the effective action of the matrix model. Although we have
found it appropriate to state the argument using the language of physics, 
it
should be stressed that it eventually involves nothing but formal
manipulations of generating functions, and thus is completely rigorous.

Unfortunately we have not found a way of analytically computing the 
generating
function of the diagrams depicted in Fig.~2 for general $n$, although a 
number
of cases ($n=1$ \STh, $n=2$ \ZJZb, $n=\infty$ \JZJb\ and $n=-2$ \PZJc)
can be solved exactly. In 
the
second part of the paper we therefore proceed to describe how the integer
coefficients of the generating function can be found, up to any desired 
order
$p$, using a computer. The number of diagrams to be enumerated scales as
$12^p$ for $p \gg 1$, but our algorithm does the counting in a most
expeditious fashion, only using a time $\sim 2.7^p$. Once again, we have
inspired ourselves from theoretical physics, reformulating the counting
process as the action of a {\it transfer matrix}, which is an object 
normally
used to describe the discrete time evolution of a quantum system.
Mathematicians may think of the transfer matrix as a linear operator that 
acts
on a set of basis states consisting of appropriately defined sub-diagrams.

{}From the description given below it will be evident that the 
applicability of
our transfer matrix method extends far beyond the context of the present
paper. Indeed, it can be generalized to the computation of generating
functions for all sorts of planar diagrams, provided that the latter have 
at
least one {\it external leg}. In particular, we are presently unable to
generate results for {\it closed} knots (usually known in the mathematics
literature simply as knots, as opposed to {\it tangles} which possess 
external
legs).

We finally state our results, here given up to a maximal order of $p=22$, 
and we analyze the asymptotic behavior of the various generating functions. 
The number of objects at order $p \gg 1$ is generally supposed to have the
asymptotic behavior $c \mu^p p^{-\alpha}$, where, in the language of
statistical physics, $\mu$ is a critical temperature and $\alpha$ a 
critical
exponent. Amazingly, the theory of two-dimensional quantum gravity 
provides a
conjecture for $\alpha$, based solely on the O($n$) symmetry of the 
underlying
matrix model, whereas the leading scaling exponent $\mu$ can only be 
accounted
for in the exactly solvable cases. Comparing the conjecture for 
$\alpha(n)$
with our numerical results, we find good agreement in its expected range 
of
validity ($0 \le n \le 2$).

%

\newsec{The matrix model and its renormalization}
The objects we want to consider are tangles with $2k$ ``external legs'',
that is roughly speaking the data 
of $k$ intervals embedded in a ball $B$ and whose endpoints are
given distinct points on the boundary $\der B$, plus an arbitrary number of
(unoriented) circles
embedded in $B$, all intertwined, and considered up to orientation
preserving homeomorphisms of $B$ that reduce to the identity on $\der B$.
Tangles with $4$ external legs will be simply called tangles.
As mentioned in the introduction,
we represent these objects using diagrams, that is regular projections
on a plane,
and restrict ourselves to alternating diagrams for
which under- and over-passes alternate as one follows any connected
component.

A natural framework is to introduce a {\it matrix model}\/ which
generates in its Feynman diagram expansion these link diagrams.
We shall define it now.

\subsec{Definition of the $O(n)$ matrix model}
As in \refs{\PZJ,\ZJZb,\PZJc},
we start with the following matrix integral over $N \times N$ hermitean matrices 
\eqn\mmm{
Z^{(N)}(n,g)=\int\! \prod_{a=1}^n \d M_a
\, \E{N\,\tr\left(-{1\over 2} \sum_{a=1}^n M_a^2+{g\over 4} \sum_{a,b=1}^n
M_a M_b M_a M_b\right)}}
where $n$ is a positive integer. 

Expanding in power series in $g$ generates
Feynman diagrams with double edges (``fat graphs'') drawn in $n$ colors
in such a way that colors cross each other at the vertices.
By taking the large $N$ limit one selects the planar diagrams,
which can be redrawn as alternating link diagrams, cf Fig.~\feya.
\fig\feya{A planar Feynman diagram of \mmm\ and the corresponding
alternating link diagram.}{\epsfxsize=8cm\epsfbox
{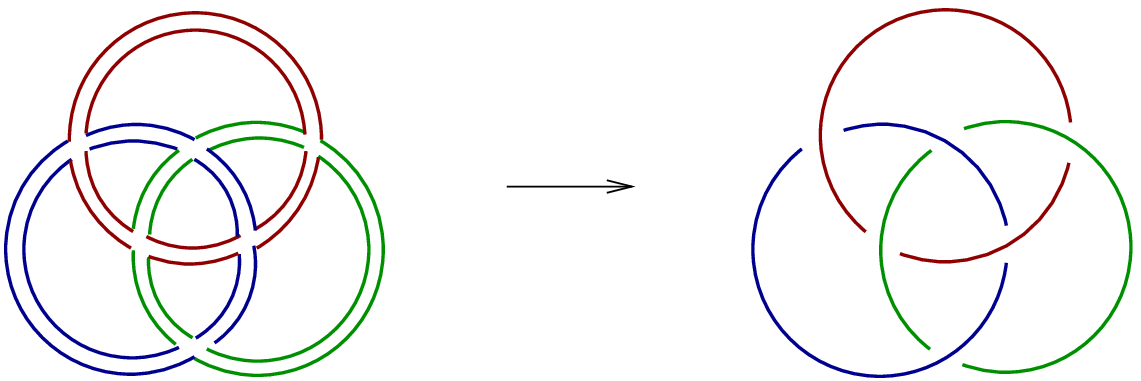}}
More precisely, the large $N$ ``free energy''
\eqn\mmmb{F(n,g)=\lim_{N\to\infty}{\log Z^{(N)}(n,g)\over N^2}}
is a double generating function of the number $f_{k;p}$ of alternating 
link diagrams with
$k$ connected components and $p$ crossings (weighted by the inverse of their
symmetry factor, and with mirror images identified):
\eqn\dblgen{F(n,g)=\sum_{k=1}^\infty\sum_{p=1}^\infty f_{k;p}\, n^k g^p}

If one is interested in counting objects with a weight of $1$,
one cannot consider the free energy
which corresponds to closed diagrams, but instead correlation
functions of the model which generate diagrams with external
legs, that is tangle diagrams.
Typically, we shall be interested in the two-point function
\eqn\two{G(n,g)\equiv \lim_{N\to\infty}\left<{1\over N} \tr M_a^2\right>
}
where the measure on the matrices $M_a$ is given by Eq.~\mmm\ 
and $a$ is any fixed index, which generates tangle diagrams with two
external legs; and the connected
four-point functions
\eqna\four
$$\eqalignno{
\Gam_1(n,g)&=\lim_{N\to\infty}\left< {1\over N}\tr (M_a M_b)^2\right>&\four{.1}\cr
\Gam_2(n,g)&=\lim_{N\to\infty}\left< {1\over N}\tr (M_a^2 M_b^2)\right>-G(n,g)^2&\four{.2}\cr
}$$
where $a$ and $b$ are two distinct indices, which generate tangle diagrams with four
external legs of type $1$ and $2$ (see Fig.~\types). 
\fig\types{Tangles of types 1 and 2.}{\epsfxsize=8cm\epsfbox{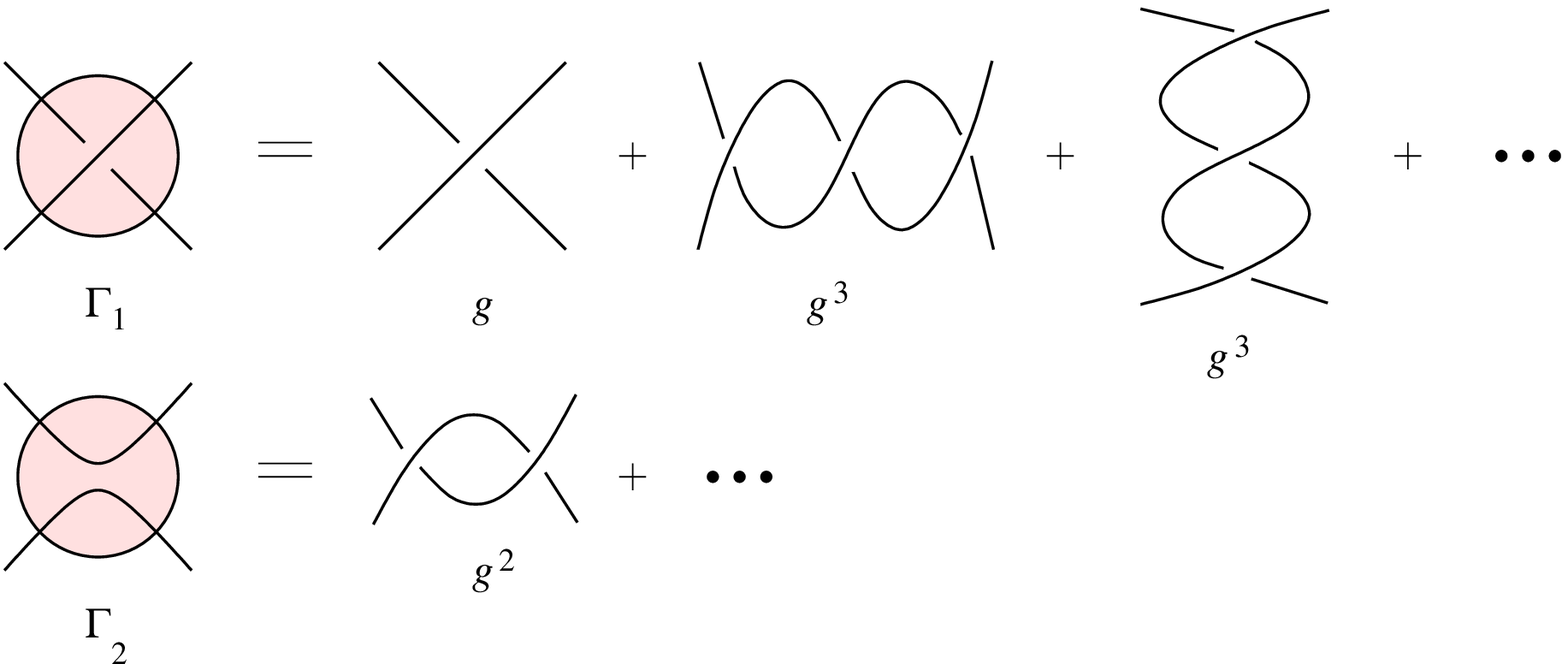}}

\subsec{Renormalization of the $O(n)$ model}
The model presented above counts tangle {\it diagrams}, but not tangles.
There is a redundancy in the counting since
to a given tangle/link corresponds many diagrams.
In order to properly count tangles, this redundancy must be removed.
In the case of alternating diagrams one can distinguish
two steps. First one must find a way to
select only {\it reduced}\/ diagrams which contain no irrelevant
crossings (Fig.~\nprime~a)); such diagrams will have minimum number of crossings.
It turns out to be convenient to introduce at this point a closely
related notion: a link is said to be {\it prime}\/ if it cannot be
decomposed into two pieces in the way depicted on Fig.~\nprime~b). It is clear
that at the level of diagrams, forbidding decompositions of the type
of Fig.~\nprime~b) automatically implies that the diagram is reduced; and
we shall therefore restrict ourselves to prime links and tangles.
\fig\nprime{a) An irrelevant crossing. b) A non-prime link.}{\epsfxsize=6cm\epsfbox{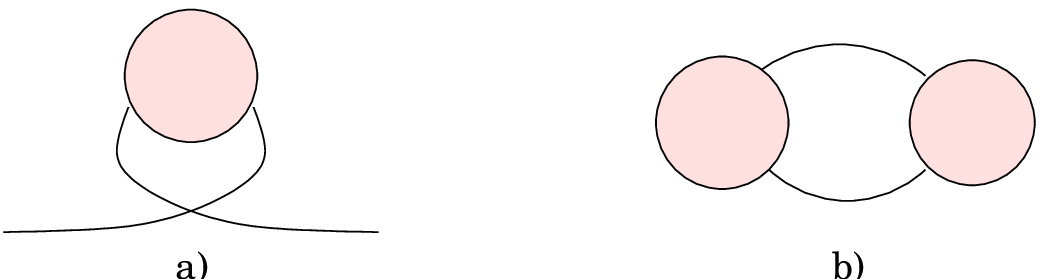}}
There may still be several reduced diagrams corresponding to the same
link: according to the flyping conjecture, proved in \MTh, two such diagrams
are related by a finite sequence of flypes, see Fig.~\flyp.
\fig\flyp{A flype.}{\epsfxsize=6cm\epsfbox{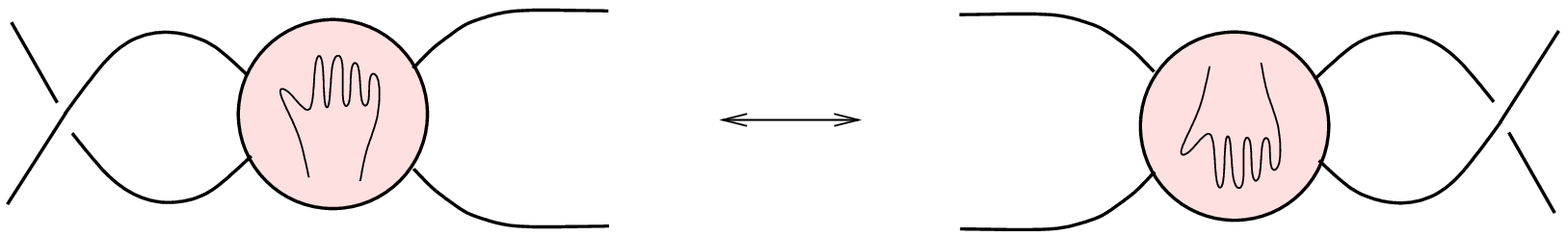}}

To summarize, there are two problems: a) the diagrams generated by applying
Feynman rules are not necessarily reduced or prime; b) several
reduced diagrams may correspond to the same knot due to the flyping
equivalence. Following the study of \PZJc, we note
(cf Figs.~\nprime\ and \flyp) that this
``overcounting'' is local in the diagrams in the sense that problem
a) is related to the existence of sub-diagrams with 2 external legs,
whereas problem b) is related to a certain class of sub-diagrams with
4 external legs. Clearly such graphs can be cancelled by the inclusion
of appropriate {\it counterterms}\/ in the action.
We are therefore led to the conclusion that we must
{\it renormalize}\/ the quadratic and quartic interactions of
\mmm. Renormalization theory tell us that we should include
in the action from the start every term compatible with the symmetries
of the model, since they will be generated dynamically by the 
renormalization. 
A key observation is that, while there is only one quadratic
$O(n)$-invariant term, there are {\it two}\/ quartic $O(n)$-invariant
terms, which leads to a generalized model with 3 coupling constants
in the action:
\eqn\mmmgen{
Z^{(N)}(n,t,g_1,g_2)=\int\! \prod_{a=1}^n \d M_a
\, \E{N\,\tr\left(-{t\over 2} \sum_{a=1}^n M_a^2
+{g_1\over 4} \sum_{a,b=1}^n (M_a M_b)^2
+{g_2\over 2} \sum_{a,b=1}^n M_a^2 M_b^2
\right)}}
The Feynman rules of this model now allow loops of different colors to ``avoid''
each other, which one can imagine as tangencies (Fig.~\feyren).
\fig\feyren{Vertices of the generalized $O(n)$ matrix model.}{%
\epsfxsize=4cm\epsfbox{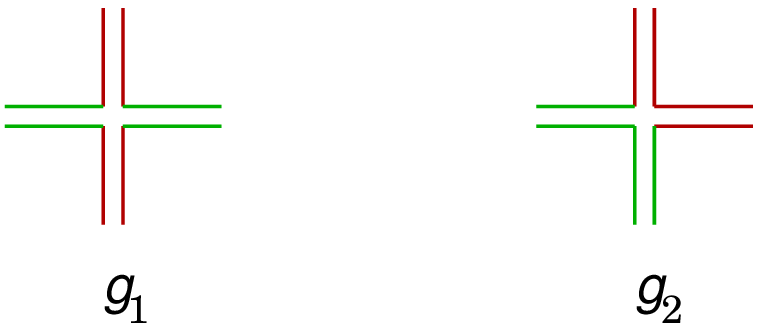}}
We define
again the correlation functions $G(n,t,g_1,g_2)$ and $\Gam_i(n,t,g_1,g_2)$
(Eqs.~\two\ and \four{}), and want to extract from them
the counting of colored alternating tangles with external legs.

The idea is to find the expressions of $t(g)$, $g_1(g)$ and
$g_2(g)$ as a function of the renormalized coupling constant $g$,
in such a way that the overcounting is suppressed and the correlation
functions are generating series in $g$ (and $n$) of the number of colored
tangles.

The derivation of the renormalization equations was performed in \PZJc.
Consider first the removal of irrelevant crossings and non-prime links.
It is clear that one must remove all two-legged
subdiagrams, that is impose
\eqn\ren{
G(n,t,g_1,g_2)=1
}
This implicitly fixes $t(g)$.

\fig\elemflyp{Breaking a flype into two elementary flypes.}{\epsfxsize=12cm\epsfbox{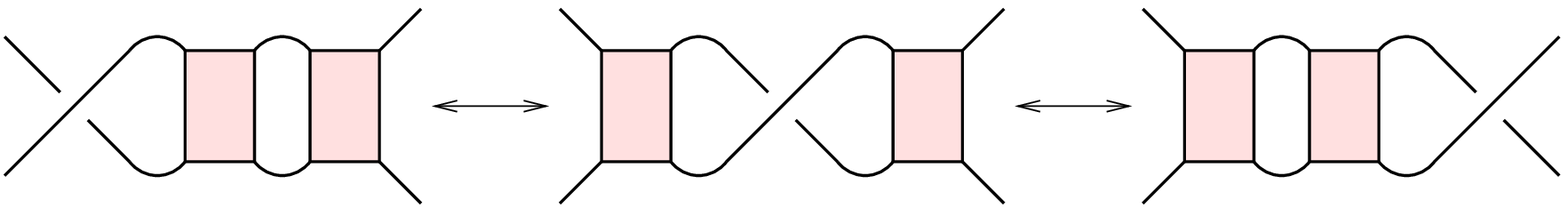}}
Next, consider the flyping equivalence. 
A flype can be made of several ``elementary'' flypes (Fig.~\elemflyp), 
an elementary flype being by definition one that cannot be decomposed 
any more in
this way. In the terminology of QFT, these elementary flypes 
are one simple vertex connected by two edges to a non-trivial 
two-particle irreducible in the horizontal channel tangle diagram (H-2PI).
We need to introduce
auxiliary generating functions $H'_1(g)$, $H'_2(g)$ and $V'_2(g)$
for non-trivial H-2PI tangles of type 1, of type 2 and of type 2 rotated by
$\pi/2$ respectively.
By considering all possible insertions of elementary
flypes as tangle sub-diagrams of a diagram,
we find (Fig.~\renflyp)
that the renormalization of $g_1$ and $g_2$ is
given by:\eqna\renc
$$\eqalignno{
g_1(g)&=g(1-2H'_2(g))&\renc{.1}\cr
g_2(g)&=-g(H'_1(g)+V'_2(g))&\renc{.2}\cr
}
$$
\fig\renflyp{Counterterms needed to cancel flypes.}{\epsfxsize=12cm\epsfbox{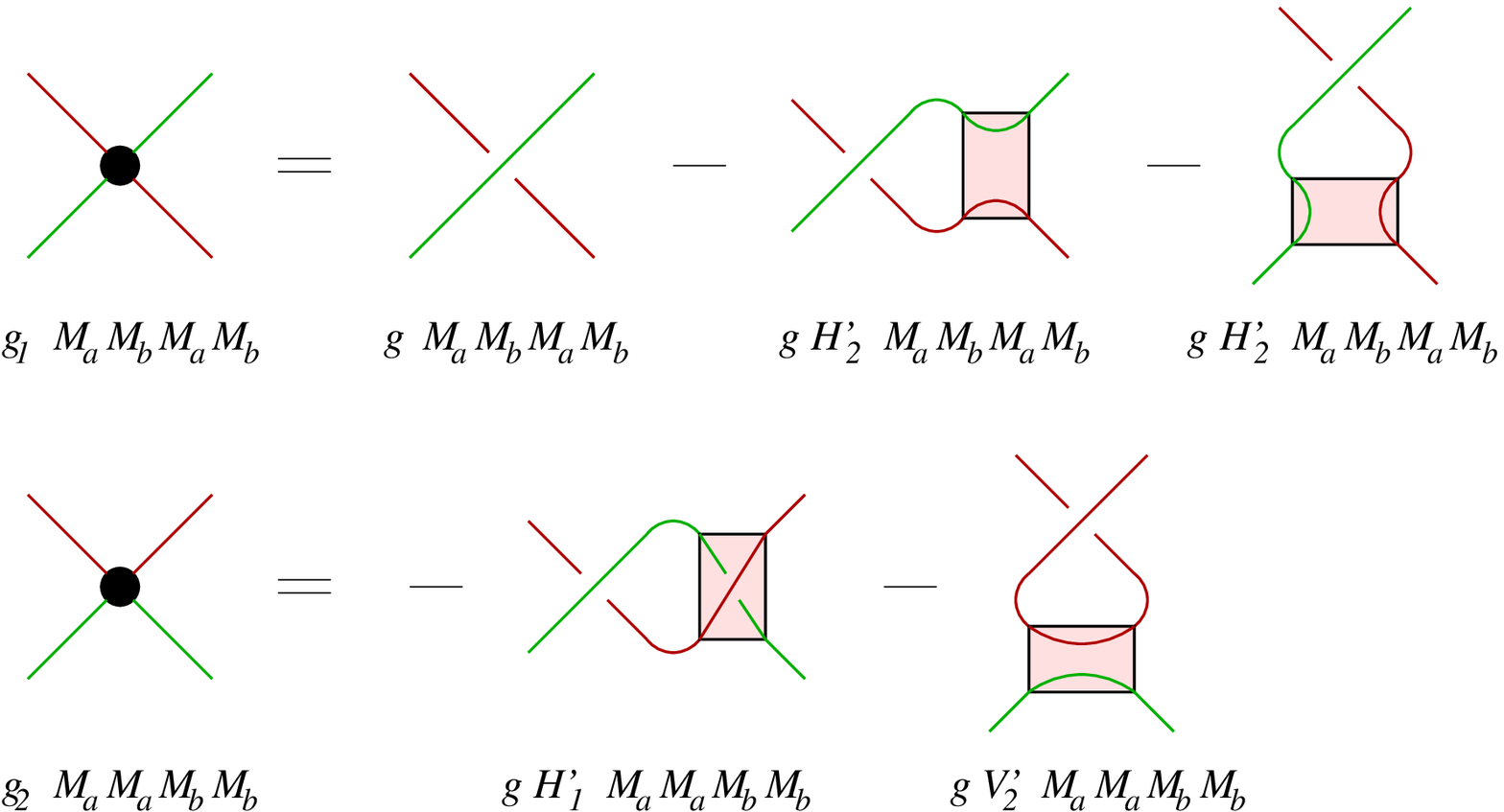}}
All that is left is to find the expressions of the auxiliary generating
functions in terms of known quantities. They are easily obtained
by decomposing the four-point functions in the horizontal and vertical 
channels \ZJZb:
\eqna\deci
$$\eqalignno{
H'_2\pm H'_1 &=1-{1\over (1\mp g)(1+\Gam_2\pm\Gam_1)}
&\deci{\rm a}\cr
H'_2+nV'_2+H'_1&=1-{1\over (1-g)(1+(n+1)\Gam_2+\Gam_1)}
&\deci{\rm b}\cr
}$$
The three renormalization equations \ren\ and \renc{}, supplemented by
the expressions \deci{}, are enough to fix $t(g)$, $g_1(g)$ and $g_2(g)$.
Solving them gives access to the $\Gam_i$,
which are the generating series of the numbers of prime alternating tangles
of type $i$.
However, we can go further.
By computing other correlation functions in the model and composing
them with the solutions $t(g)$, $g_1(g)$, $g_2(g)$ of the equations above,
one can extract
the generating functions of the number of alternating tangles with an
arbitrary number of external legs. An example will be given below.

\newsec{An algorithm for the counting of decorated planar diagrams}
Unfortunately the matrix model described in section 2.1 cannot
be solved for a generic value of $n$ (see however \refs{\ZJZ,\ZJZb,\PZJc}
for solutions of particular values). 
We therefore turn to the description of an algorithm
that allows to enumerate the required diagrams to a given order $p$
of crossings. This is one of several similar
algorithms described in \refs{\JZJ,\JZJb}.
The numbers of diagrams thus computed are the input needed
for the combinatorial treatment of section 2.2. 

The principle of the algorithm is to iterate a transfer matrix which
``builds'' the diagrams slice by slice:
starting from an initial state consisting of all external legs 
(two in the case under consideration),
the system is time evolved through the addition of $p$ intersections,
until an empty final state is obtained. 

We shall first concentrate on the enumeration of (prime,
alternating) tangle {\it diagrams}, which correspond to
the ``unrenormalized'' model of section 2.
Adding tangencies,
which is needed to take into account the flyping equivalence, will be
discussed in Section 3.2, since it is an elementary extension of the
algorithm.

\subsec{The single-step algorithm} 
\fig\singlestep{Working principle of the single-step algorithm.
a) A two-legged knot diagram with $p=6$ intersections and $k=3$
connected components. The edges are labelled from A to M.
b) The same diagram in the time-slice representation.
For reasons of clarity, the time slices are not drawn in chronological
order.}
{\epsfxsize=14cm\epsfbox{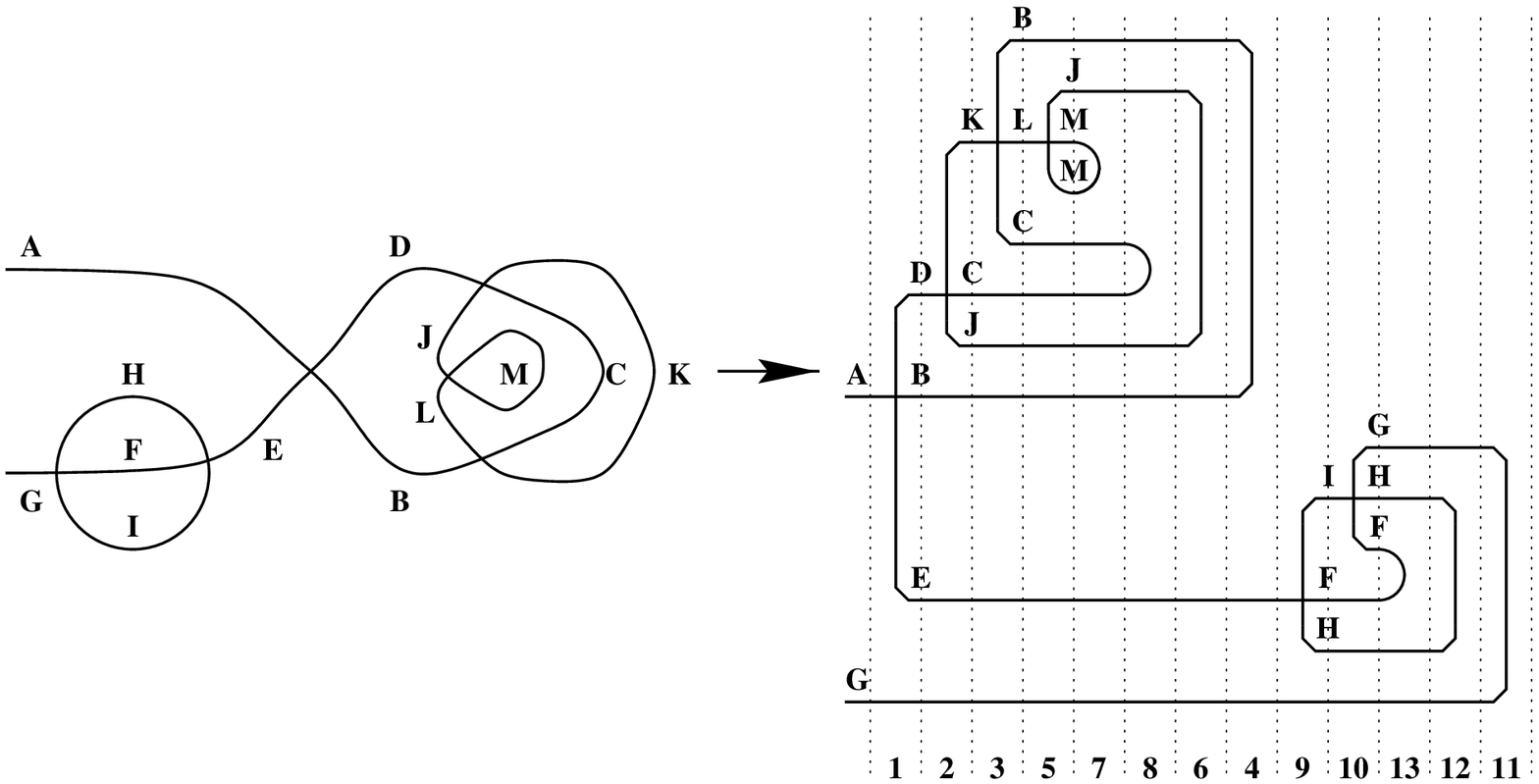}}

The algorithm we want to describe, called ``single-step algorithm''
in \JZJb, is best explained by an example.
Let us consider the tangle
diagram shown in Fig.~\singlestep.
Let us start from an initial state given by the two external
legs (edges A and G). We want to ``evolve'' the various open lines
by following them and adding the new crossings they meet.

Moving along either of the edges A or G, a new line segment
(DE resp.~HI) is encountered. The question then arises which of
these to process first. We resolve this ambiguity by stipulating
that {\it in any given state, we evolve the line which at that
instant is uppermost}.
At time $t=1$, the edge A thus becomes B, and the new line segment
DE is added. The edge D is now the new top line.
Analogously, at the instants $t=2$ and $t=3$, the top line (D resp.\ K)
crosses a new line segment, which is then added to the current state.
We can formalize this by stating the transformation rule shown
in Fig.~\transform.1.

At time $t=3$, the new top line carries the label B, which
was however already produced by the transformation acting at $t=1$.
We therefore proceed, at $t=4$, to the identification of the two ``copies''
of B, joining them through an arch. This is an example of the general
transformation rule shown in Fig.~\transform.2. The addition of an
arch means that the lines intermediate between the two instances of B
(at positions $p=1$ and $q$ on Fig.~\transform.2) can henceforth not
communicate with the lines at the exterior of the arch. These
``trapped lines'' must therefore eventually evolve to the empty state
(vacuum), independently of the rest of the diagram. 
Since the current top line must always be treated first, the
evolution of a possible set of ``trapped lines'' must take place at
a later time. Thus,
on Fig.~\singlestep, we cannot always {\it draw} the time-slices in chronological
order. 
The existence of a number of trapped lines is visualized on
Fig.~\transform.2 by a {\it delimiter} (shown as a gray rectangle),
which separated the remaining lines into two {\it blocks}. Lines in
different blocks cannot communicate, and so the transformation 2)
only applies when $p$ and $q$ belong to the same block.

\fig\transform{The two types of transformations in the single-step algorithm.
1) Addition of a new line segment. 2) Identification of the top line
(at position $p$) with another line (at position $q$), accompanied with
the creation of a new block.}
{\epsfxsize=7cm\epsfbox{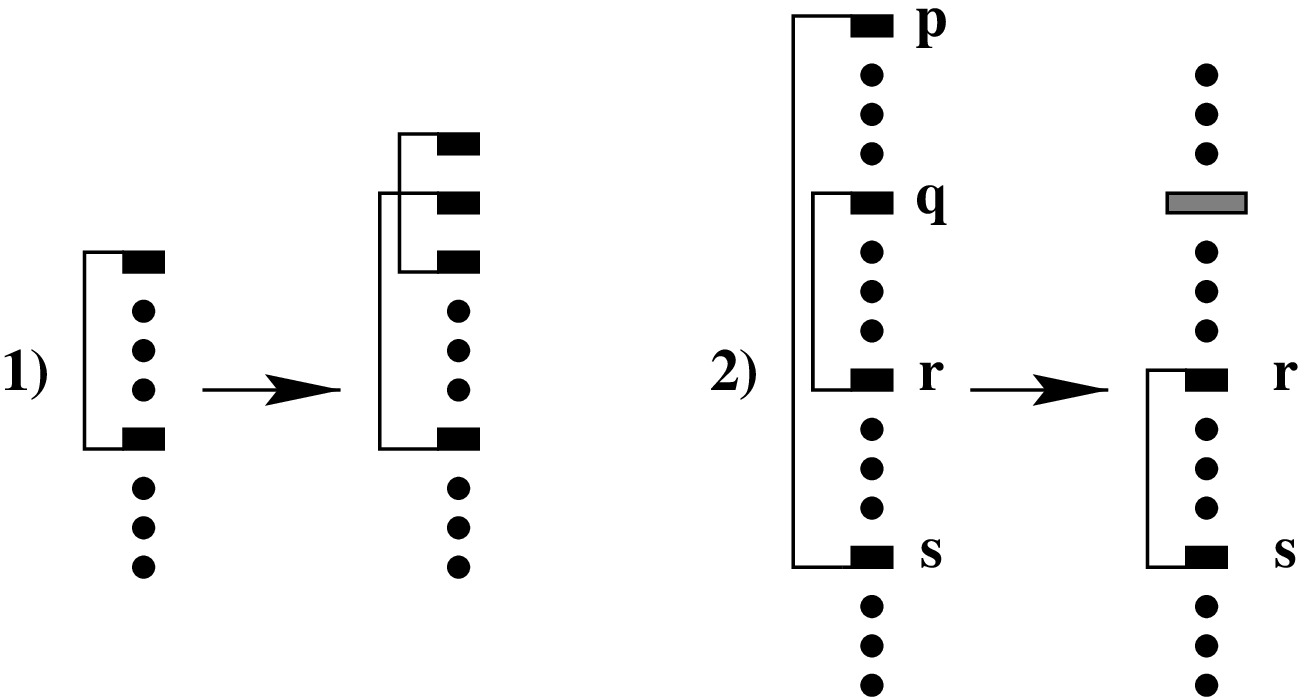}}

The purpose of the transfer matrix is not only to count the total number
of tangle diagrams, but to do so for any fixed number of connected components.
In particular, when performing a type 2 transformation, we need to know
whether the points $p$ and $q$ were already connected though an arbitrary
number of edges at an {\it earlier} time. On Fig.~\transform\ we have
represented this information by a number of lines on the left, connecting
the points at a given instant into pairs. It may thus happen that on
Fig.~\transform.2, $r=p$ and $s=q$. In this case, the type 2 transformation
marks the completion of one connected component in the tangle diagram.

\fig\intermed{Intermediate states produced by applying the single-step
algorithm to the tangle diagram shown in Fig.~\singlestep.}
{\epsfxsize=10cm\epsfbox{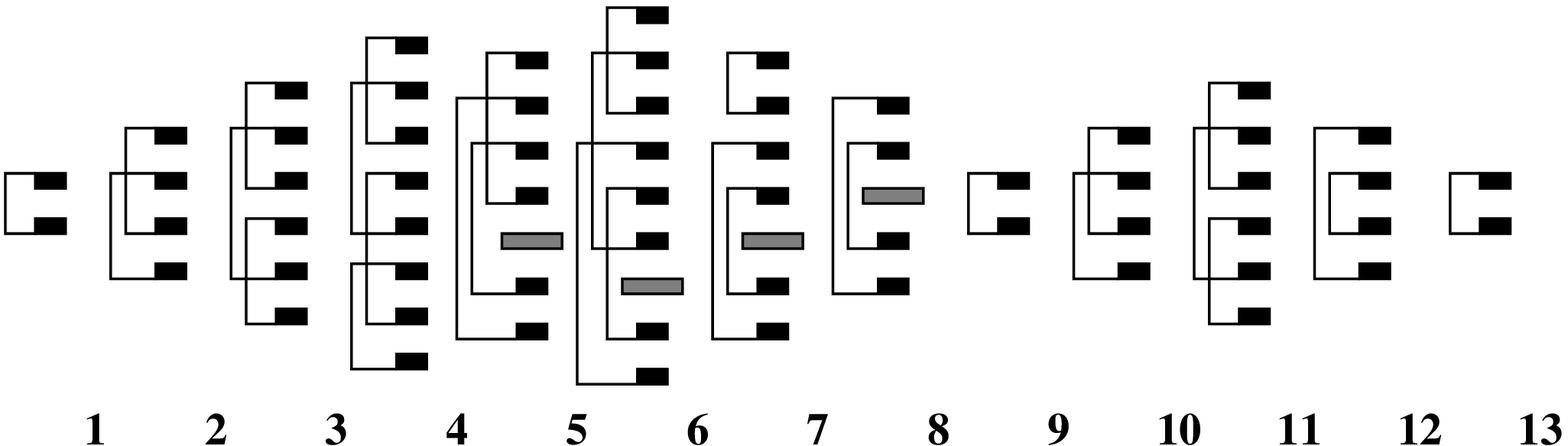}}

We are now ready to define the set of {\it states} on which the transfer
matrix acts. A state is defined by an even number of points (represented
on Fig.~\transform\ as black rectangles), connected into pairs by means
of the edges encountered at previous times. In addition, the points
are divided into $\ell+1$ blocks by means of $\ell \ge 0$ delimiters.
On Fig.~\intermed\ we show the set of intermediate states corresponding
to the time-slice representation of the tangle of Fig.~\singlestep.

Finally, we must define the transfer matrix $T$ which counts {\it all}
tangle diagrams with $\tilde{p}$ vertices and $k$ connected components.
Its entries $T_{ab}$, where $a$ and $b$ are two basis states of the
kind just defined, are $0$ unless $b$ is a descendant of $a$.
An allowed state $b$ is a descendant of $a$ if it can be obtained
via a transformation of one of the two types shown on Fig.~\transform\ 
(for an arbitrary even $q\ge 2$ belonging to the same block as $p=1$),
followed by an arbitrary number of simplifications. $T_{ab}$ is
then the sum over all transformations from $a$ to $b$ of
the corresponding weight: $1$ or $n$ depending on whether
one closes a connected component or not.
The simplications are elimination of
superfluous delimiters, and reduction of states to a ``normal form''
by using various symmetries of the transfer matrix \JZJb.

\subsec{Tangencies}
Until now we have been discussing the enumeration of tangle diagrams
in which every vertex represents a crossing. However, to account for
the flype equivalence we need to enumerate more general diagrams with
$p_1$ intersections and $p_2$ tangencies, as discussed in Section 2.
Adding a tangency rather than an intersection is obtained by modifying
the transformation if Fig.~\transform.1, so that the two points added
at time $t+1$ are both immediately above (resp.~immediately below)
the uppermost point at time $t$.
any given time, specifies how many tangency transformations
were used prior to that instant. The desired diagrams are then generated
by sequences of $p_1$ crossing transformations (type 1), 
$p_2$ trangency transformations,
and $p_1+p_2+1$ transformations of type 2, so that no intermediate state
is empty.

\newsec{Numerical results and analysis}
The implementation of the above algorithm
provides an efficient
numerical way to count the diagrams needed for our purposes. 
We present here some of the data obtained in \refs{\JZJ,\JZJb}.

First, we give one example of the ``raw data'' obtained from the algorithm.
It corresponds to counting a certain set of diagrams without taking into
account topological equivalences: in the present case, {\it connected} 
two-legged alternating
diagrams (called two-legged knot diagrams
in \JZJ), see Tab.~\resdiag. The $2$PI diagrams are precisely the reduced diagrams
of prime tangles.
The importance of going to orders as high as possible comes from the issue of the asymptotic behavior which will be discussed below.
\tab\resdiag{Total number $G$ of two-legged diagrams,
as well as the subset $\Sigma_2$ of $2$PI diagrams.}%
{\vbox{\offinterlineskip \halign{\strut\hfil$#$\hfil\quad&\vrule#&&\quad\hfil$#$\hfil\crcr p&&G&\Sigma_2\cr \omit&height2pt\cr \noalign{\hrule} \omit&height2pt\cr 0&&1&0\cr 1&&2&2\cr 2&&8&0\cr 3&&42&2\cr 4&&260&4\cr 5&&1796&12\cr 6&&13396&60\cr 7&&105706&
226\cr 8&&870772&1076\cr 9&&7420836&5156\cr 10&&65004584&24984\cr 
11&&582521748&128548\cr
12&&5320936416&663040\cr 13&&
49402687392&3514968\cr 14&&465189744448&18918792\cr 15&&4434492302426&103123906\cr 16&&42731740126228&569877652\cr 17&&415736458808868&3180066004\cr 18&&4079436831493480&17921451960\cr 19&&40338413922226212&101842206548\cr 20&&401652846850965808&583109887
600\cr 21&&4024556509468827432&3361640932872 \cr
22&&40558226664529024000&19499226668816\cr
}}} 
Let us also give two examples of renormalized data. 
Tab.~\resknot\ shows the number of four-legged prime alternating tangles of the 
two possible types (Fig.~\types) with arbitrary number of connected components, 
while Tab.~\ressix\ shows the number of six-legged tangles, 
which exist in five types (Fig.~\typesix).
\tab\resknot{Table of the number of prime alternating tangles with $4$ external legs.}{\vbox{\offinterlineskip \halign{\strut#&\enskip\vrule#\enskip &\hfil#\ \hfil&\hfil#\ \hfil&\hfil#\ \hfil&\hfil#\ \hfil&\hfil#\ \hfil&\hfil#\ \hfil&\hfil#\ \hfil &\enskip\vrule#\enskip &\hfil#\ \hfil&\hfil#\ \hfil&\hfil#\ \hfil&\hfil#\ \hfil&\hfil#\ \hfil&\hfil#\ \hfil&\hfil#\ \hfil \crcr &&&&$\Gam_1$&&&&&&&&$\Gam_2$\cr ${}_{p}{}^{k}$&&0&1&2&3&4&5&6&&0&1&2&3&4&5&6\cr \omit&height2pt&&&&&&&&height2pt\cr
\noalign{\hrule}
\omit&height2pt&&&&&&&&height2pt\cr
1&&1&&&&&&&&0\cr
2&&0&&&&&&&&1\cr
3&&2&&&&&&&&1\cr
4&&2&&&&&&&&3&1\cr
5&&6&3&&&&&&&9&1\cr
6&&30&2&&&&&&&21&11&1\cr
7&&62&40&2&&&&&&101&32&1\cr
8&&382&106&2&&&&&&346&153&24&1\cr
9&&1338&548&83&2&&&&&1576&747&68&1\cr
10&&6216&2968&194&2&&&&&7040&3162&562&43&1\cr
11&&29656&11966&2160&124&2&&&&31556&17188&2671&121&1\cr
12&&131316&71422&9554&316&2&&&&153916&80490&15295&1484&69&1\cr
13&&669138&328376&58985&5189&184&2&&&724758&425381&87865&6991&194&1\cr
14&&3156172&1796974&347038&22454&478&2&&&3610768&2176099&471620&52231&3280&103&1\cr
15&&16032652&9298054&1864884&193658&10428&260&2&&17853814&11376072&2768255&308697&15431&290&1\cr
}
}}
\font\seven=cmr7
\tab\ressix{Table of the number of prime alternating tangles with $6$ 
external legs.}{\vbox{\offinterlineskip
\halign{\strut#&\enskip\vrule#\enskip
&\hfil{\seven #\ }\hfil&\hfil{\seven #\ }\hfil&\hfil{\seven #\ }\hfil&\hfil{\seven #\ }\hfil
&\enskip\vrule#\enskip
&\hfil{\seven #\ }\hfil&\hfil{\seven #\ }\hfil&\hfil{\seven #\ }\hfil&\hfil{\seven #\ }\hfil
&\enskip\vrule#\enskip
&\hfil{\seven #\ }\hfil&\hfil{\seven #\ }\hfil&\hfil{\seven #\ }\hfil&\hfil{\seven #\ }\hfil&\hfil{\seven #\ }\hfil
&\enskip\vrule#\enskip
&\hfil{\seven #\ }\hfil&\hfil{\seven #\ }\hfil&\hfil{\seven #\ }\hfil&\hfil{\seven #\ }\hfil
&\enskip\vrule#\enskip
&\hfil{\seven #\ }\hfil&\hfil{\seven #\ }\hfil&\hfil{\seven #\ }\hfil&\hfil{\seven #\ }\hfil
\crcr
&&&$\Xi_1$&&&&&$\Xi_2$&&&&&$\Xi_3$&&&&&&$\Xi_4$&&&&&$\Xi_5$\cr
${}_{p}{}^{k}$&&0&1&2&3&&0&1&2&3&&0&1&2&3&4&&0&1&2&3&&0&1&2&3
\cr
\omit&height2pt&&&&&height2pt&&&&&height2pt&&&&&&height2pt&&&&&height2pt\cr
\noalign{\hrule}
\omit&height2pt&&&&&height2pt&&&&&height2pt&&&&&&height2pt&&&&&height2pt\cr
2&&0&&&&&1&&&&&0&&&&&&0&&&&&0&&&\cr
3&&2&&&&&0&&&&&2&&&&&&0&&&&&0&&&\cr
4&&0&&&&&7&&&&&2&&&&&&4&&&&&3&&&\cr
5&&18&&&&&6&&&&&16&2&&&&&8&&&&&9&&&\cr
6&&18&&&&&53&8&&&&42&2&&&&&42&7&&&&41&7&&\cr
7&&156&24&&&&154&6&&&&171&44&2&&&&156&14&&&&168&21&&\cr
8&&516&18&&&&609&181&6&&&748&114&2&&&&608&153&10&&&663&165&12&\cr
9&&2016&598&18&&&2956&422&6&&&2877&858&81&2&&&2850&586&20&&&3072&740&36&\cr
10&&10608&1428&18&&&11203&3498&318&6&&14037&3752&213&2&&&11918&3445&364&13&&13347&3966&438&18\cr
11&&40428&12318&1062&18&&57664&15330&738&6&&61028&19757&2511&131&2&&57602&17558&1406&26&&63393&20994&2040&54\cr
}}}
\fig\typesix{The five types of tangles with 6 external legs.}
{\epsfxsize=10cm\epsfbox{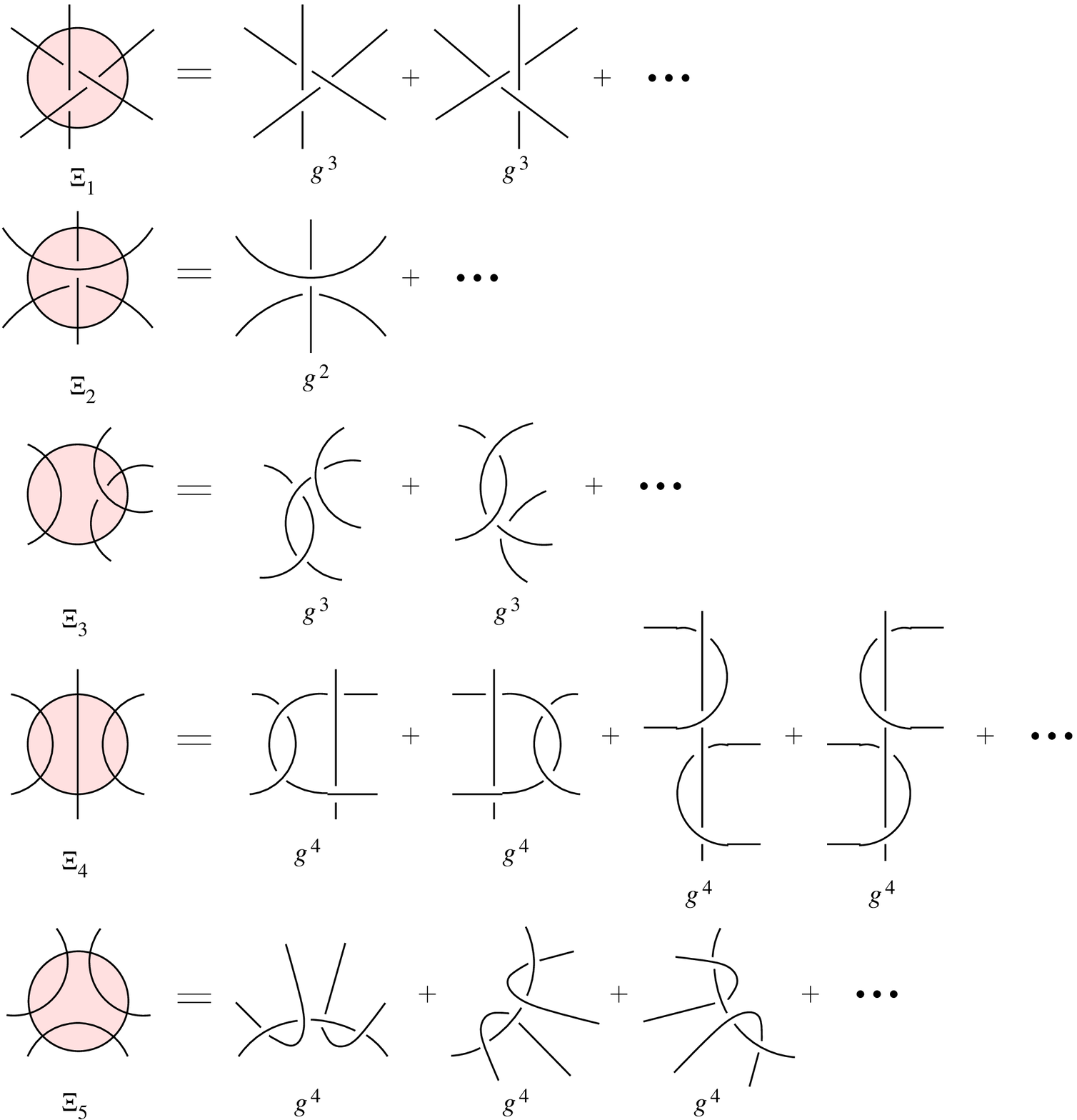}}

Finally, let us discuss the expected asymptotic behavior of these series.
Let us consider an unrenormalized generating series for tangle diagrams,
say $G(n,g)=\sum_{p=1}^\infty a_p(n) g^p$. Then it is expected that
\eqn\asya{
a_p(n){\buildrel p\to\infty\over\approx} \e{\hat{s}(n) p} p^{-\alpha(n)}
}
The conjecture holds for any correlation function in the model; for the
free energy $F(n,g)=\sum_{p=1}^\infty f_p(n) g^p$, that is the generating
series for link diagrams (the symmetry factors being most likely irrelevant
for the asymptotic behavior), it is
\eqn\asyb{
f_p(n){\buildrel p\to\infty\over\approx} \e{\hat{s}(n) p} p^{-\alpha(n)-1}
}
The bulk entropy $\hat{s}(n)$ can be easily extracted with a good accuracy
from the numerical data. However it is {\it non universal}, which implies that
it is not preserved by the renormalization of the model. On the other hand,
the exponent
$\alpha(n)$ is universal, and presumably preserved by the renormalization
(unless an unusual phase transition takes place in the model). Since the raw
data are ``cleaner'' than the renormalized data, they are better suited to
the extraction of $\alpha(n)$. A conjecture made in \PZJ, based in particular
on the KPZ relation \KPZ, is that for $n$
analytically continued to $0\le n\le 2$, one has
\eqn\expconj{
\alpha(n=-2\cos(\pi\nu))=1+1/\nu\qquad 1/2\le\nu\le1
}

Using Tab.~\resdiag\ one can try to evaluate $\alpha=\alpha(n=0)$,
the exponent of tangles with minimum number of components, which also gives
the exponent of knots according to Eq.~\asyb. The conjecture \expconj\ yields
$\alpha=3$. Unfortunately,
with the current data, it is difficult to estimate $\alpha$
without any knowledge of the subleading corrections. Indeed, a direct fit gives
$\alpha\approx 2.76$ but with very weak accuracy. In \JZJ, it was suggested
to fit the data with $a_p=\e{\hat{s}p} p^{-\alpha} (a\log p+b+o(1))$,
the result being in good agreement with the conjecture:
\eqn\fitgen{
\e{\hat{s}}=11.416\pm 0.005 \quad
\alpha=2.97\pm 0.06 \quad
a=0.04\pm0.02 \quad
b=0.1\pm0.03}
However there is little evidence for such a logarithmic correction, and the
issue remains open. A more detailed discussion of $\alpha(n)$ for all $n$
can be found in \JZJb.

\listrefs
\bye